\pdfoutput=1
\documentclass{article}
\usepackage{spconf,amsmath,graphicx}

\usepackage[absolute,showboxes]{textpos}

\TPMargin{5pt}

\newcommand{\copyrightline}{
    \begin{textblock}{0.84}(0.08,0.93)    
         \noindent
         \footnotesize
         \copyright 2020 IEEE. Personal use of this material is permitted. Permission from IEEE must be obtained for all other uses, in any current or future media, including reprinting/republishing this material for advertising or promotional purposes, creating new collective works, for resale or redistribution to servers or lists, or reuse of any copyrighted component of this work in other works.
    \end{textblock}
}

\DeclareMathOperator{\sinc}{sinc} 
\DeclareMathOperator{\rect}{rect}
\DeclareMathOperator{\abs}{abs}

\usepackage[dvipsnames]{xcolor}
\newcommand{\new}[1] {{\color{black} #1}} 

\title{Small-Footprint Keyword Spotting on Raw Audio Data with Sinc-Convolutions}

\name{Simon Mittermaier$^{\star \dagger}$ \qquad Ludwig K\"urzinger$^{\star}$ \qquad Bernd Waschneck$^{\dagger}$ \qquad Gerhard Rigoll$^{\star}$}

\address{$^{\star}$ Technische Universit\"at M\"unchen,
    Munich, Germany \\ 
    $^{\dagger}$ Infineon Technologies AG,
	Munich \& Dresden, Germany}

\begin{document}
%
\maketitle
\copyrightline
\begin{abstract}
Keyword Spotting (KWS) enables speech-based user interaction on smart devices.
Always-on and battery-powered application scenarios for smart devices put constraints on hardware resources and power consumption, while also demanding high accuracy as well as real-time capability.
Previous architectures first extracted acoustic features and then applied a neural network to classify keyword probabilities, optimizing towards memory footprint and execution time.

Compared to previous publications, we took additional steps to reduce power and memory consumption without reducing classification accuracy.
Power-consuming audio preprocessing and data transfer steps are eliminated by directly classifying from raw audio.
For this, our end-to-end architecture extracts spectral features using parametrized Sinc-convolutions.
Its memory footprint is further reduced by grouping depthwise separable convolutions.
Our network achieves the competitive accuracy of $96.4\%$ on Google's Speech Commands test set with only $62$k parameters. 
\end{abstract}
\section{Introduction}
\label{sec:intro}


Speech processing enables natural communication with smart phones or smart home assistants, e.g., Amazon Echo, Google Home. 
However, continuously performing speech recognition is not energy-efficient and would drain batteries of smart devices. 
Instead, most speech recognition systems passively listen for utterances of certain wake words such as ``Ok Google", ``Hey Siri", ``Alexa", etc. to trigger the continuous speech recognition system on demand.
This task is referred to as keyword spotting (KWS) or wake-word detection.
There are also uses of KWS where a view simple speech commands (e.g. ``on", ``off") are enough to interact with a smart device. 

Conventional hybrid approaches to KWS first divide their audio signal in time frames to extract features, e.g., Mel Frequency Cepstral Coefficients (MFCC).
A neural net then estimates phoneme or state posteriors of the keyword Hidden Markov Model in order to calculate the keyword probability using a Viterbi search.
\new{The wake-word is then recognized when the keyword probability reaches a predefined threshold.}
In recent years, end-to-end architectures gained traction that directly classify keyword posterior probabilities based on the previously extracted features, e.g., ~\cite{NN_KWS, HelloEdge, ResNet, TCResNet, fernandez2007application}. 

Typical application scenarios imply that the device is powered by a battery, and possesses restricted hardware resources to reduce costs.
Therefore previous works optimized towards memory footprint and operations per second.
In contrast to this, we tune our neural network towards energy conservation in microcontrollers motivated by obervations on power consumption, as detailed in Sec.~\ref{ssec:requirements}.
To extract meaningful and representative features from raw audio, our architecture uses parametrized Sinc-convolutions (SincConv) from SincNet~\cite{SincNet1}.
We use Depthwise Separable Convolutions (DSConv)~\cite{DSConvforTranslation, Xception} that preserve time-context information while at the same time compare features in different channels. 
To further reduce the number of network parameters, which is key for energy efficiency, we group DSConv-layers, a technique we refer to as Grouped DSConv (GDSConv). 

Our key contributions are:
\begin{itemize}
	\item We propose a neural network architecture tuned towards energy efficiency in microcontrollers grounded on the observation that memory access is costly, while computation is cheap~\cite{MemVsOps}. 
	\item Our keyword-spotting network classifies on raw audio employing SincConvs while at the same time reducing the number of parameters using (G)DSConvs. 
	\item Our base model with $122$k parameters performs with the state-of-the-art accuracy of $96.6\%$ on the test set of Google’s Speech Commands dataset, on par with TC-ResNet~\cite{TCResNet} that has $305$k parameters and requires separate preprocessing. Our low-parameter model achieves $96.4\%$ with only $62$k parameters.
\end{itemize}

\section{Related Work}
\label{sec:related}

Recently, CNNs have been successfully applied to KWS~\cite{HelloEdge,ResNet,TCResNet}.
Zhang et al. evaluated different neural network architectures (such as CNNs, LSTMs, GRUs) in terms of accuracy, computational operations and memory footprint as well as their deployment on embedded hardware \cite{HelloEdge}.
They achieved their best results using a CNN with DSConvs.
Tang et al. explored the use of Deep Residual Networks with dilated convolutions to achieve a high accuracy of $95.8\%$ \cite{ResNet}, while keeping the number of parameters comparable to \cite{HelloEdge}.
Choi et al. build on this work as they also use a ResNet-inspired architecture.
Instead of using 2D convolution over a time-frequency representation of the data they convolve along the time dimension and treat the frequency dimension as channels \cite{TCResNet}.

This bears similarities with our approach as we are using 1D convolution along the time dimension as well.
However, all the approaches mentioned classify from MFCCs or similar preprocessed features.
Our architecture works directly on raw audio signals.
\new{Previous work has been done with using neural networks for directly modeling raw audio for wake word detection in a hybrid approach \cite{RawAudioWWAmazon}.
There is, however, a recent trend in the speech domain towards using CNNs on raw audio data directly in an end-to-end approach \cite{SincNet1, SincNet3,  FacebookRaw1, FacebookRaw2}.}
Ravanelli et al. present an effective method of processing raw audio with CNNs, called SincNet.
Kernels of the first convolutional layer are restricted to only learn shapes of parametrized sinc functions.
This method was first introduced for Speaker Recognition \cite{SincNet1} and later also used for Phoneme Recognition \cite{SincNet3}. To the best of our knowledge, we are the first to apply this method to the task of KWS.

\new{The first convolutional layer of our model is inspired by SincNet and we combine it with DSConv that has first been introduced in the domain of Image Processing \cite{Xception, Mobilnets} and applied to KWS in~\cite{HelloEdge} and neural machine translation \cite{DSConvforTranslation}.}
\new{Kaiser et al.~\cite{DSConvforTranslation} also introduce the ``super-separable'' convolution, a DSConv that further reduces the number of parameters using grouping.}
The idea of Grouped Convolutions was first used in AlexNet \cite{AlexNet} to reduce parameters and operations and to enable distributed computing of the model over multiple GPUs. 
We denominate the combination of grouping and DSonv as GDSConv in our work and use it for our smallest model.



\section{Model}
\label{sec:model}


\subsection{Keyword-Spotting on Battery-Powered Devices}
\label{ssec:requirements}

Typical application scenarios for smart devices imply that the device is powered by a battery, and possesses restricted hardware resources.
The requirements for a KWS system in these scenarios are (1) very low power consumption to maximize battery life, (2) real-time or near real-time capability, (3) low memory footprint and (4) high accuracy to avoid random activations and to ensure responsiveness.

Regarding real-time capability, our model is designed to operate on a single-core microcontroller capable of 50 MOps per second~\cite{HelloEdge}.
We assume that in microcontrollers the memory consumption of a KWS neural network is associated with its power consumption:
Reading memory values contributes most to power consumption which makes re-use of weights favorable.
While in general large memory modules leak more power than small memory modules,
one read operation from RAM costs far more energy than the corresponding multiply-and-accumulate computation~\cite{horowitz20141, MemVsOps}.
In addition to the parameter-reducing approach in this work, further steps may be employed to reduce power consumption such as quantization, model compression or optimization strategies regarding dataflows that depend on the utilized hardware platform~\cite{horowitz20141, MemVsOps, chen2016eyeriss, shahnawaz2018studying}.

\subsection{Feature Extraction using SincConvs}
\label{ssec:sincconv}

SincNet~\cite{SincNet1} classifies on raw audio by restricting the filters of the first convolutional layer of a CNN to only learn parametrized sinc functions, i.e., $\sinc⁡(x)=\sin(x)/x$.
One sinc function in the time domain represents a rectangular function in the spectral domain,
therefore two sinc functions can be combined to an ideal band-pass filter:
\begin{align}
    g[n, f_1, f_2 ] &=& 2 f_2  \sinc ⁡(2\pi f_2 n )- 2 f_1  \sinc⁡ (2\pi f_1 n) \\
    G[f, f_1, f_2 ] &=& \rect ⁡(\frac{f}{2f_2})  - \rect ⁡(\frac{f}{2f_1})
\end{align}
Performing convolution with such a filter extracts the parts of the input signal that lie within a certain frequency range.
SincNet combines Sinc-convolutions with CNNs;
as we only use the feature extraction layer of this architecture, we label this layer as SincConv to establish a distinction to SincNet.

\begin{figure}[htb]
\begin{minipage}[b]{\linewidth}
  \centering
  \centerline{\includegraphics{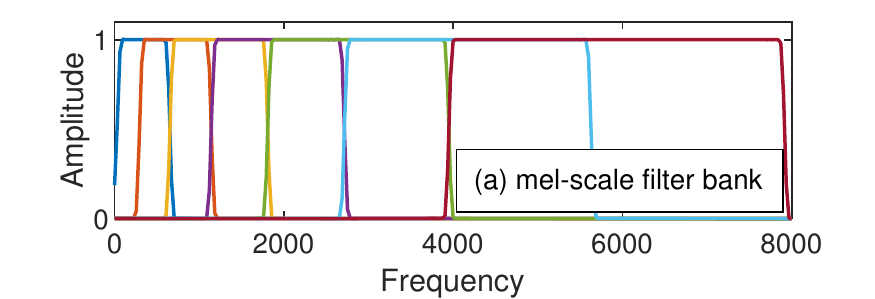}}
\end{minipage}
\begin{minipage}[b]{\linewidth}
  \centering
  \centerline{\includegraphics{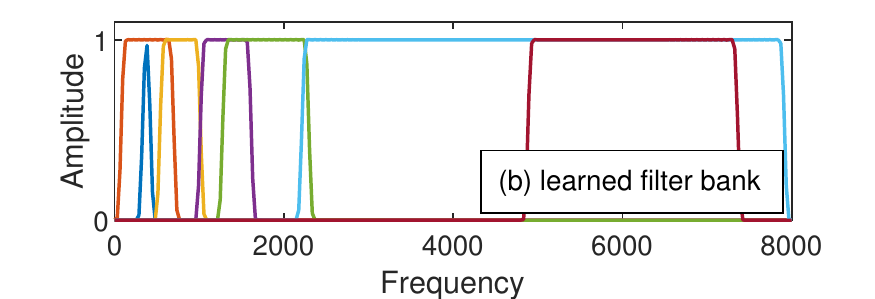}}
\end{minipage}
\caption{SincConv filter bank with 7 filters, (a) after initialization with mel-scale weights and (b) after training on~\cite{SpeechCommands}.}
\label{fig:Filterbank}
\end{figure}

\begin{figure}[htb]
\begin{minipage}[b]{.49\linewidth}
  \centering
  \centerline{\includegraphics{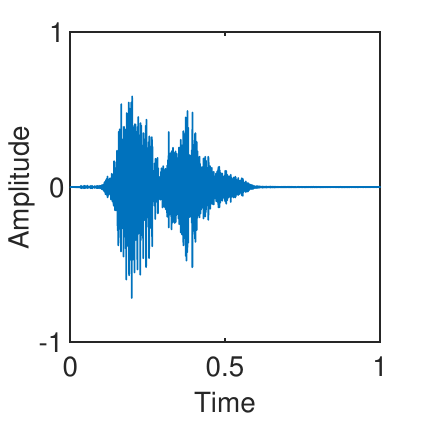}}
\end{minipage}
\hfill
\begin{minipage}[b]{0.49\linewidth}
  \centering
  \centerline{\includegraphics{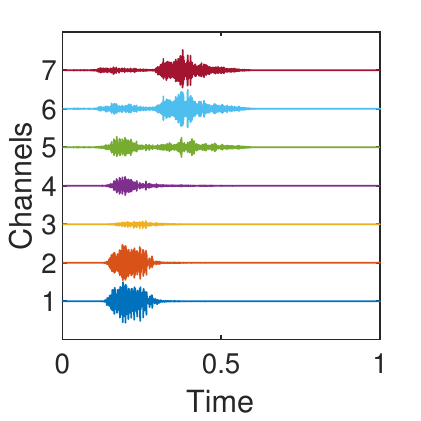}}
\end{minipage}
\caption{An audio sample of the keyword ``yes'' is convolved with the 7-filter SincConv layer from Fig.~\ref{fig:Filterbank} to extract meaningful features.}
\label{fig:SincConvOutput}
\end{figure}

Compared to one filter of a regular CNN, were the number of parameters is derived from its kernel width, e.g., $k=100$~\cite{SincNet1},
Sinc-convolutions only require two parameters to derive each filter, the lower and upper cut-off frequencies ($f_1,f_2$), resulting in a small memory footprint.
SincConv filters are initialized with the cutoff frequencies of the mel-scale filter bank and then further adjusted during training.
Fig.~\ref{fig:Filterbank} visualizes this adjustment from initialization to after training.
SincConv filter banks can be easily interpreted, as the two learned parameter correspond to a specific frequency band.
Fig.~\ref{fig:SincConvOutput} visualizes how a SincConv layer with 7 filters processes an audio sample containing the word ``yes''.


\subsection{Low-Parameter GDSConv Layers}
\label{ssec:gdsconv}

\begin{figure}[htb]
\centering
\includegraphics[width=\linewidth]{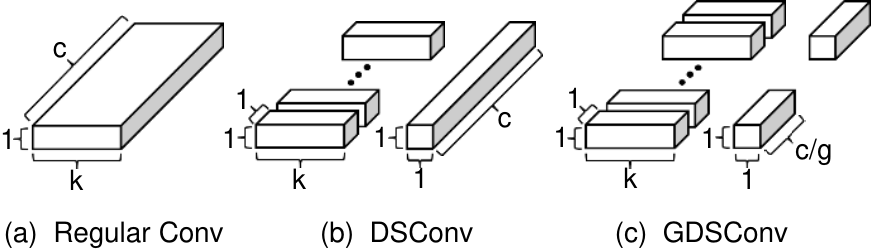}
\caption{Steps from a regular convolution to the grouped depthwise separable convolution.
(a) Regular 1D convolutional layers perform convolution along the time axis, across all channels. 
(b) The DSConv convolves all channels separately along the time dimension (depthwise), and then adds a 1x1 convolution (i.e., a pointwise convolution) to combine information across channels. 
(c) 
The GDSConv is performed by partitioning the channels into $g$ groups and then applying a DSConv on each group.
Our base model employs DSConv layers, and our low-parameter model GDSConv layers.
}
\label{fig:GDSConv}
\end{figure}

DSConv have been successfully applied to the domain of computer vision \cite{Xception, Mobilnets}, neural translation \cite{DSConvforTranslation} and KWS~\cite{HelloEdge}.
Fig.~\ref{fig:GDSConv} provides an overview of the steps from a regular convolution to the GDSConv. \new{A mathematical description is given in \cite{DSConvforTranslation}.}

The number of parameters of one DSConv layer amounts to $N_{\text{DSConv}}=k\cdot c_{in}+c_{in}\cdot c_{out}$ with the kernel size $k$ and the number of input and output channels $c_{in}$ and $c_{out}$ respectively;
the first summand is determined by the depthwise convolution, the second summand by the pointwise convolution~\cite{DSConvforTranslation}.
In our model configuration, the depthwise convolution only accounts for roughly $5\%$ of parameters in this layer, the pointwise for $95\%$.
We therefore reduced the parameters of the pointwise convolution using grouping by a factor $g$ to $N_{\text{GDSConv}}=k\cdot c_{in}+\frac{c_{in}\cdot c_{out}}{g}$,
rather than the parameters in the depthwise convolution.
To allow information exchange between groups we alternate the number of groups per layer, namely 2 and 3, as proposed in~\cite{DSConvforTranslation}.

\subsection{Two Low-Parameter Architectures}
\label{ssec:architecture}

\begin{figure}[htb]
\centering
\includegraphics[width=\linewidth]{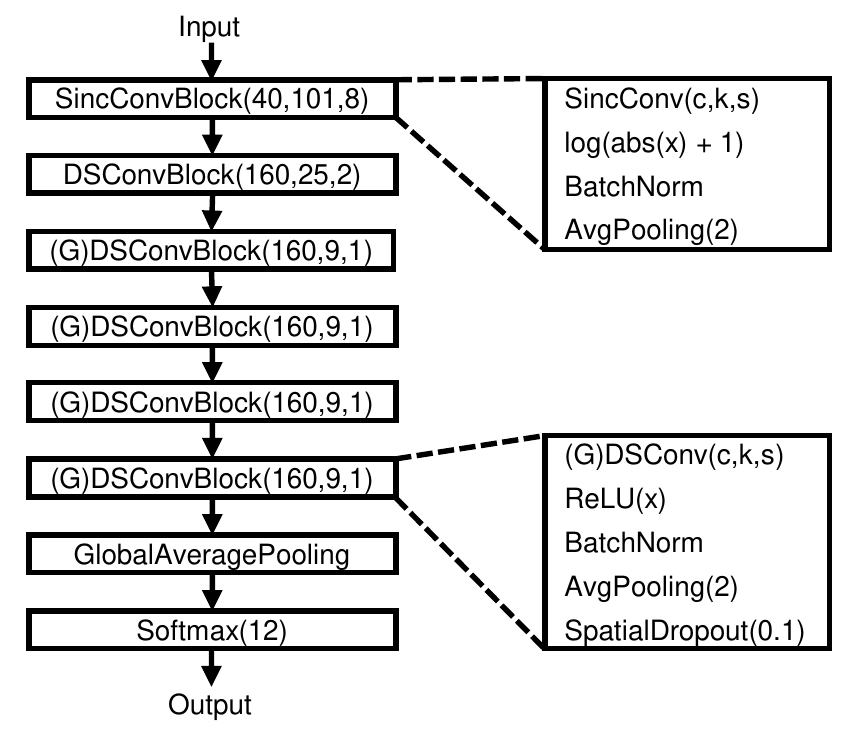}
\caption{The model architecture as described in Sec.~\ref{ssec:architecture}. Parameter configurations of convolutions are given as $c,k,s$ that represent the number of output channels, kernel length and stride, respectively.
In our low-parameter model, convolutions are grouped from the third to the sixth layer.
}
\label{fig:Model}
\end{figure}

The SincConv as the first layer extracts features from the raw input samples, as shown in Fig. \ref{fig:Model}. 
As non-linearity after the SincConv we opt to use log-compression, i.e., $y=\log(\abs(x)+1)$, instead of a common activation function (e.g., ReLU).
This has also shown to be effective in other CNN architectures for raw audio processing~\cite{FacebookRaw1, FacebookRaw2}.
Five (G)DSConv layers are then used to process the features further:
The first layer has a larger kernel size and scales the number of channels to $160$.
The other four layers have each $160$ input and output channels.
Each (G)DSConv block contains the (G)DSConv layer, batch normalization~\cite{BatchNorm} and spatial dropout~\cite{SpatialDropout} for regularization, as well as average pooling to reduce temporal resolution.
After the (G)DSConv blocks, we use global average pooling to receive a $160$-element vector that can be transformed to class posteriors using a Softmax layer to classify $12$ classes, i.e., $10$ keywords as well as a class for \emph{unknown} and for \emph{silence}.

The low-parameter model is obtained by grouping the DSConv layers with an alternating number of groups between 2 and 3.
For the configuration shown in Fig.~\ref{fig:Model}, the base model has $122$k parameters.
After grouping, the number of parameters is reduced to a total of $62$k.
\new{Hyperparameters have been selected through extensive experimentation based on the best validation accuracy for a given number of parameters comparable to models in ~\cite{HelloEdge,TCResNet}.
}

\section{Evaluation}
\label{sec:evaluation}

\subsection{Training on the Speech Commands Dataset}

We train and evaluate our model using Google's Speech Commands data set~\cite{SpeechCommands}, an established dataset for benchmarking KWS systems.
The first version of the data set consists of $65$k one-second long utterances of $30$ different keywords spoken by $1881$ different speakers. 
The most common setup consists of a classification of 12 classes: ``yes", ``no", ``up", ``down", ``left", ``right", ``on", ``off", ``stop", ``go", \emph{unknown}, or \emph{silence}.
The remaining $20$ keywords are labeled as \emph{unknown}, samples of provided background noise files as \emph{silence}.
To ensure the benchmark reproducibility, a separate test set was released with a predefined list of samples for the \emph{unknown} and the \emph{silence} class.
The second version of the dataset contains $105$k samples and five additional keywords~\cite{SpeechCommands}.
However, previous publications on KWS reported only results on the first version, therefore we focused on the first version and additionally report testing results on version 2 of the dataset.

Every sample from the training set is used in training, this leads to a class imbalance as there are much more samples for \emph{unknown}.
Class weights in the training phase assign a lower weight to samples labeled as \emph{unknown} such that the impact on the model is proportional to the other classes. 
This way, the model can see more \emph{unknown} word samples during training without getting biased.
Our model is trained for 60 epochs with the Adam optimizer~\cite{Adam} with an initial learning rate of 0.001 and learning rate decay of 0.5 after 10 epochs;
the model with the highest validation accuracy is saved to evaluate accuracy on the test set.

\subsection{Results and Discussion}

The base model composed of DSConv layers without grouping achieves the state-of-the-art accuracy of 96.6\% on the Speech Commands test set.
The low-parameter model with GDSConv achieves almost the same accuracy of 96.4\% with only about half the parameters.
This validates the effectiveness of GDSConv for model size reduction. 

Table~\ref{table:1} lists these results in comparison with related work.
Compared to the DSConv network in~\cite{HelloEdge}, our network is more efficient in terms of accuracy for a given parameter count.
Their biggest model has a 1.2\% lower accuracy than our base model while having about 4 times the parameters.
Choi et al.~\cite{TCResNet} has the most competitive results while we are still able to improve upon their accuracy for a given number of parameters.
They are using 1D convolution along the time dimension as well which may be evidence that this yields better performance for audio processing or at least KWS.

As opposed to previous works, our architecture does not use preprocessing to extract features,
but is able to extract features from raw audio samples with the SincConv layer.
That makes it possible to execute a full inference as floating point operations,
without requiring additional hardware modules to process or transfer preprocessed features.
Furthermore, we deliberately opted to not use residual connections in our network architecture,
considering the memory overhead and added difficulty for hardware acceleration modules.

For future comparability, we also trained and evaluated our model on the newer version 2 of the Speech Commands data set; see Table~\ref{table:2} for results.
On a side note, we observed that models trained on version 2 of the Speech Commands dataset tend to perform better on both the test set for version 2 and the test set for version 1~\cite{SpeechCommands}.

\begin{table}[tb]
\centering
\begin{tabular}{lll}
\hline
Model                            & Accuracy & Parameters\\
\hline \hline
DS-CNN-S \cite{HelloEdge}        & $94.1\%$ & $39$k       \\
DS-CNN-M \cite{HelloEdge}        & $94.9\%$ & $189$k      \\
DS-CNN-L \cite{HelloEdge}        & $95.4\%$ & $498$k      \\
ResNet15 \cite{ResNet}           & $95.8\%$ & $240$k      \\
TC-ResNet8 \cite{TCResNet}      & $96.1\%$ & $66$k       \\
TC-ResNet14  \cite{TCResNet}    & $96.2\%$ & $137$k      \\
TC-ResNet14-1.5 \cite{TCResNet} & $\mathbf{96.6\%}$ & $305$k      \\
\hline
\textbf{SincConv+DSConv}         & $\mathbf{96.6\%}$  & $122$k      \\
\textbf{SincConv+GDSConv}        & $96.4\%$  & $\mathbf{62}$\textbf{k}       \\
\hline
\end{tabular}
\caption{Comparison of results on the Speech Commands dataset~\cite{SpeechCommands}.}
\label{table:1}
\end{table}

\begin{table}[tb]
\centering
\begin{tabular}{lll}
\hline
Model                     & Accuracy & Parameters \\
\hline \hline
\textbf{SincConv+DSConv}  & $97.4\%$  & $122$k       \\
\textbf{SincConv+GDSConv} & $97.3\%$  & $62$k        \\
\hline
\end{tabular}
\caption{Results on Speech Commands version 2~\cite{SpeechCommands}.}
\label{table:2}
\end{table}

\section{Conclusion}
\label{sec:conclusion}


Always-on, battery-powered devices running keyword spotting require energy efficient neural networks with high accuracy.
For this, we identified the parameter count in a neural network as a main contributor to power consumption,
as memory accesses contribute far more to power consumption than the computation.
Based on this observation, we proposed an energy efficient KWS neural network architecture by combining feature extraction using SincConvs with GDSConv layers.

Starting with the base model composed of DSConvs that have already less parameters than a regular convolution, we achieved state-of-the-art accuracy on Google's Speech Commands dataset.
We further reduce the number of parameters by grouping the convolutional channels to GDSConv, resulting in a low-parameter model with only $62$k parameters.

\vfill\pagebreak

\bibliographystyle{IEEEbib}
\bibliography{refs}

\end{document}